\documentclass[draft,12pt,grl]{agutex}

\usepackage{times}
\usepackage{amsbsy}

\usepackage{amsmath}
\usepackage{amsthm}
\usepackage{eucal}
\usepackage{amssymb}
\usepackage{mathrsfs}

\usepackage[dvips]{graphicx}

\setkeys{Gin}{draft=false}

\authorrunninghead{MEYER-VERNET ET AL.}

\titlerunninghead{Dust observed by \textit{Wind}}

\authoraddr{N. Meyer-Vernet,
LESIA - CNRS - Observatoire de Paris, 5 place Jules Janssen, 92195 Meudon, France.
(nicole.meyer@obspm.fr)}

\begin{document}

\title{The importance of monopole antennas for dust observations:\\ why \textit{Wind}/WAVES does not detect nanodust
}

%
%


\authors{
N. Meyer-Vernet\altaffilmark{1},
M. Moncuquet\altaffilmark{1},
K. Issautier\altaffilmark{1},
and A.~Lecacheux\altaffilmark{1}
}

\altaffiltext{1}
{LESIA, CNRS, Observatoire de Paris, Meudon, France, UPMC, Universit\'{e} Paris Diderot.}
%
%


\begin{abstract}

The charge released by impact ionization of fast dust grains impinging on spacecraft is at the basis of a well-known technique for dust detection by wave instruments. Since most of the impact charges are recollected by the spacecraft, monopole antennas  generally detect a much greater signal than dipoles. This is illustrated by comparing dust signals in monopole and dipole mode on different spacecraft and environments. It explains the weak sensitivity of \textit{Wind}/WAVES dipole antennas for dust detection, so that  it is not surprising that this instrument did not detect the interplanetary nanodust discovered by STEREO/WAVES. We propose an interpretation of the \textit{Wind} dust data, elsewhere discussed by \citet{mal14}, which explains the observed pulse amplitude and polarity for interstellar dust impacts, as well as the non-detection of nanodust.  This proposed mechanism might be the dominant dust detection mechanism by some wave instruments using dipole antennas.

\end{abstract}

%
%

%

\begin{article}

\section{Introduction}

\begin{figure}[h]
\center
\noindent\includegraphics[width=20pc]{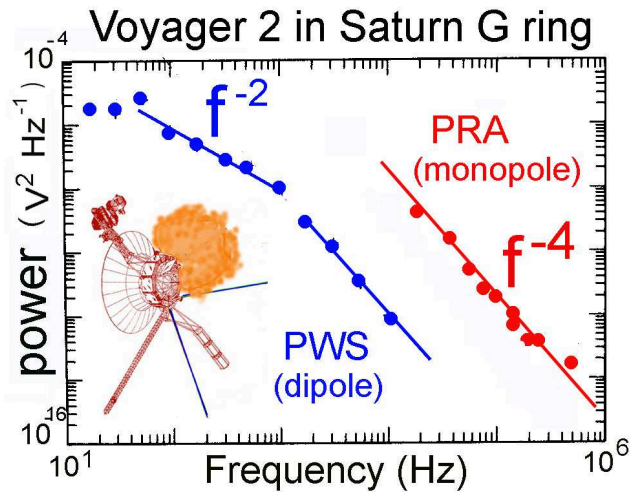}
\caption{Voltage power spectrum measured in Saturn's G ring by the (high frequency) radio (PRA) and (lower frequency) plasma wave (PWS) instruments on Voyager 2, with respectively monopole
and dipole antennas. At similar frequencies, the power is higher by nearly four orders of magnitude on the monopole than on the dipole. Adapted from \cite{man11} with kind permission from Springer Science and Business Media.}  \label{voyager}
\end{figure}

When a dust grain impacts a spacecraft at high speed, the electric charges produced by impact ionization induce an  electric pulse which can be detected by onboard wave instruments. This technique was pioneered when the Voyager spacecraft crossed Saturn ring plane and the onboard radio (PRA)  \citep{aub83} and plasma wave (PWS) \citep{gur83} instruments detected micron-sized dust grains. These   instruments detected signals of very different amplitudes  because they had different wave capture capabilities:  PRA used the antennas as monopoles whereas PWS used them as dipoles (Figure \ref{voyager}) - a difference which must be taken into account in analyzing the signals. 

This serendipitous discovery opened the way to a new technique which was later  used to measure micron-sized grains on Voyager in the dust rings of Uranus \citep{mey86a,gur87} and Neptune \citep{gur91,ped91}, in cometary environments on Vega  (e.g. \citep{obe96}) and ISEE-3/ICE \citep{gur86,mey86b}, in Saturn's E ring with Cassini (e.g. \citep{kur06,mon13}), in the solar wind near 1 AU with STEREO \citep{zas12,bel12}, and even in the outer solar system \citep{gur97}. The technique was recently extended to measure fast nanodust in Jovian nanodust streams with Cassini \citep{mey09b}, and led to the discovery by STEREO  of interplanetary nanodust picked-up by the solar wind  \citep{mey09a,zas12,lec13}.

\citet{mal14} have recently discussed the voltage pulses produced on \textit{Wind}/WAVES by dust impacts. The non-detection by this instrument of the fast nanodust discovered by STEREO/WAVES  led these authors to suggest an inconsistency between the STEREO and \textit{Wind} results on nanodust, arguing that both spacecraft were in close proximity and had similar electric field waveform capture capabilities.

We submit that this discrepancy between \textit{Wind} and STEREO is not surprising. In addition to the very dissimilar antenna geometries, both instruments have completely different electric field waveform capture capabilities: indeed, STEREO can use the antennas as monopoles whereas \textit{Wind} only uses them as dipoles. Since this difference between dipoles and monopoles is sometimes ignored outside the radio astronomy community, producing  confusion as in the case of the recent paper by \cite{mal14}, it is worth discussing it first. We  then propose an interpretation of the \textit{Wind} dust data, which may be relevant with other wave instruments using dipole antennas.

\section{Dipole versus monopole electric antennas}

Since the spacecraft surface area generally exceeds that of the electric antennas by several orders of magnitude, most of the dust impacts occur on the spacecraft, which recollects most of the impact charges of sign opposite to its  floating potential.

In dense planetary magnetospheres, the ambient plasma generally dominates the charging currents, producing a negative floating potential, whereas in the solar wind photoelectron emission dominates, producing a positive floating potential (e.g. \cite{man14}). Therefore, in dense plasmas the target tends to recollect the positively charged ions of the impact plasma, whereas in the solar wind the target recollects the electrons. This produces a potential pulse on the target of the same sign as the recollected charges.

This mechanism produces a potential pulse $\delta V_{sc} \sim Q/C_{sc}$ on a spacecraft of capacitance $C_{sc}$ recollecting the charge $Q$, so that each monopole antenna, say $y$ (detecting the potential between itself and the spacecraft structure)  measures a pulse $\delta V_y \simeq -\delta V_{sc}$. In contrast, a dipole antenna (detecting the potential between two  arms, say $y_+$ and   $y_-$) measures via this mechanism a much smaller signal, produced by circuit imbalances (common mode rejection factor, differences in antenna and base capacitances).

Figure \ref{voyager} shows the power spectra detected by the (high frequency) radio \citep{aub83} and by the (low frequency) plasma wave instrument \citep{gur83} onboard Voyager 2 in Saturn's G ring, with  monopole and dipole antennas, respectively. The power spectrum is proportional to the square of the pulse voltage amplitude produced by individual impacts and to the impact rate (see \citep{mey85} for a detailed calculation of the spectral shape). One sees that the dipole antennas recorded a power  smaller than the monopole by nearly four orders of magnitude, in agreement with the  signal being mainly produced by recollection by the spacecraft of the impact charges produced by grain impacts  on its surface  \citep{mey85,obe96,mey96,mey98}. The smaller power spectrum  observed in dipole mode can be due in particular to recollection by an antenna of  a fraction of the  charges produced by impacts on the spacecraft or on itself  \citep{gur83}.

\begin{figure}[h]
\center
\noindent\includegraphics[width=30pc]{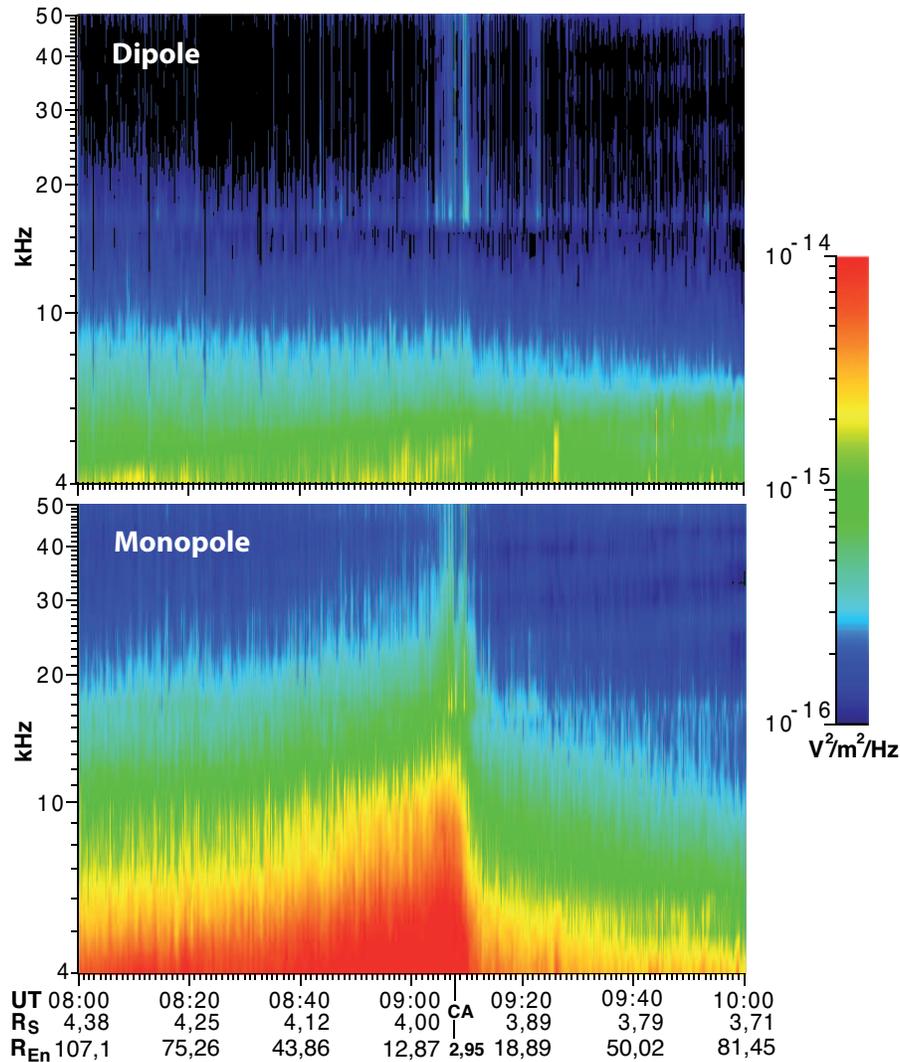}
\caption{Time-frequency electric power spectral density measured by Cassini/RPWS between 4 and 50 kHz during the first close approach of Enceladus (on 2005/03/09) with dipole (top) and monopole (bottom) antennas. Both spectrograms are calibrated in $V^2/$m$^2/$Hz (color chart). The signal produced by E ring dust impacts is clearly seen in monopole mode near the closest approach (CA) of Enceladus orbit. The time and Cassini's distance from Saturn (in R$_S = 60330$ km) and from Enceladus (in R$_{En} = 252$ km) are indicated every 20 min at the bottom.} \label{moncuquet}
\end{figure}

Figure \ref{moncuquet} compares the electric power spectra measured by the Cassini/RPWS high frequency receiver  \citep{gur04} simultaneously in dipole (top) and monopole (bottom) mode during the first close approach to Enceladus in 2005. The increase in power density  due to micron-sized dust impacts in the E-ring \citep{kur06} is clearly visible on the monopole antenna, whereas the dipole only measures the weaker plasma quasi-thermal and impact noise \citep{sch13}. This property was recently used  by \citet{mon13} to determine the large scale structure of the E ring.

\begin{figure}[h]
\center
\noindent\includegraphics[width=30pc]{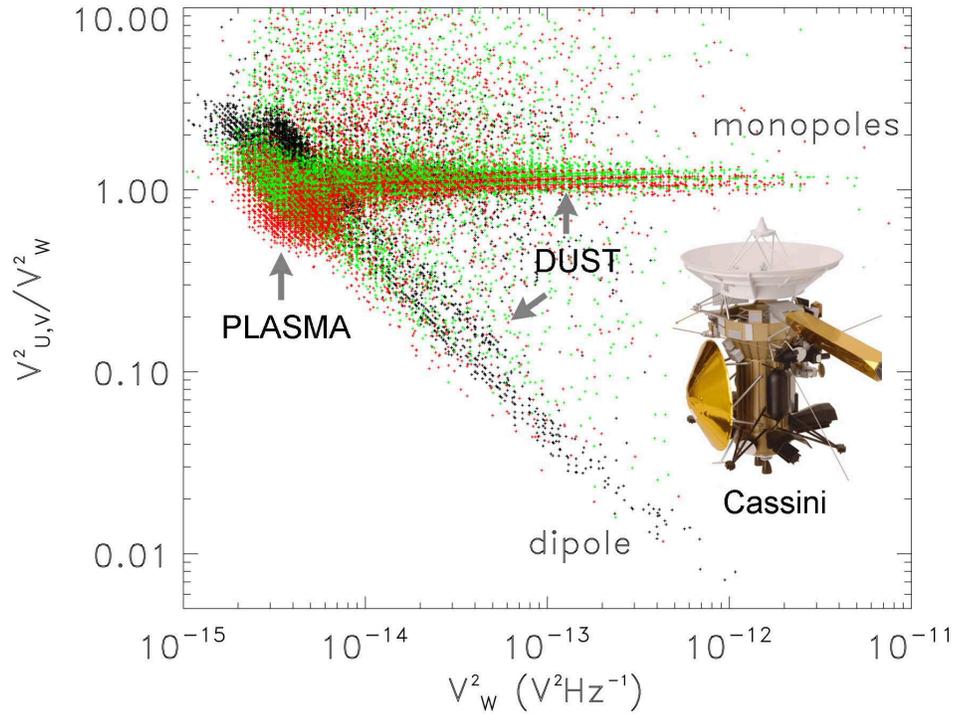}
\caption{Cassini/RPWS high-frequency receiver data in the Jovian outer magnetosheath. Ratios of the power on two monopoles (red and green) to that on the other one as a function of the latter, and ratio of the power on the dipole to that on the monopole (black). The dust impacts yield similar signals on each monopole, whereas the dipole records mainly the plasma thermal noise (of smaller amplitude). Adapted from \cite{mey09b}.}  \label{cassini}
\end{figure}

Figure \ref{cassini} compares the power spectra measured by Cassini/RPWS  in dipole and monopole mode near Jupiter, when this instrument detected Jovian nanodust streams \citep{mey09b} simultaneously to their measurements by the Cosmic Dust Analyzer \citep{gra01}. One sees that the large power produced by dust impacts is only observed by the monopoles,  with similar amplitudes on the three  monopoles, the small difference being consistent with the differences in antenna capacitances and receiver gains. In contrast, the dipole essentially  measures the weaker  plasma quasi-thermal and impact shot noise calculated by  \citet{mey89}.

On Voyager and Cassini, the electric antennas have a radius of about 1 cm, and a surface area smaller than the spacecraft one by nearly two orders of magnitude. In contrast, the \textit{Wind}/WAVES antennas  are thin wires of 0.19 mm radius \citep{sit97} extending perpendicular to the cylindrical spacecraft surface. Therefore, they are expected to collect a still smaller fraction of the impact released  charges.

Could the voltage observed by \textit{Wind}/WAVES be nevertheless produced by recollection by the antennas of the impact-produced charges, as is often assumed to explain the dust observations in dipole mode? This explanation, originally invoked by \citet{mal14}, is inconsistent with the data since the floating potential of surfaces in the solar wind is positive, making them recollect electrons, rather than (positive) ions; hence the recollection (of electrons) by an antenna arm should produce on it a negative potential pulse instead of a positive one. The voltage observed in dipole mode should then be of sign opposite to that required to yield  the correct direction of interstellar dust flow.

\section{Alternative detection mechanism with thin dipole antennas}

Therefore, impact charge recollection by the antennas cannot  explain the voltage pulses measured by \textit{Wind}/WAVES. The mechanism of destabilisation of the photoelectrons surrounding the antennas suggested by \citet{pan12,pan13} and implemented by \citet{zas12} on STEREO requires antennas of large radius and an adequate geometry (the STEREO/WAVES antennas extend close to the spacecraft plane faces). Even if the geometry were adequate, the 0.19 mm radius of the \textit{Wind}  spin-plane dipoles would preclude this mechanism to be significant because the photoelectron current is too small \citep{pan12}. What is then the  origin of the pulses detected by the \textit{Wind} dipole antennas?

We propose that  they are produced by the electrostatic voltage  induced on the antennas by the impact produced positive ions after the spacecraft has recollected the electrons. This mechanism is consistent with the voltage sign observed on \textit{Wind} for interstellar grain impacts, since the antenna arm closer to the impact site will then measure a larger positive voltage. Let us estimate the amplitude. The voltage produced on each antenna arm by a charge $Q$   can be calculated by averaging along the length of this  antenna arm  the Debye shielded Coulomb potential at distance $r$, $Q\times  e^{-r/L_D}/(4\pi \epsilon _0 r)$, $L_D$ being the Debye length; this holds for charges of speed much smaller than the electron thermal speed, which is the case for the impact produced ions. Therefore, a  cloud of electric charge $Q$ produces a voltage pulse of amplitude  $\delta V \sim \alpha Q/(4\pi \epsilon _0 L)$ (in order of magnitude) on an antenna arm of physical length $L$ when it is closer to the antenna axis than the shielding Debye length (see \citep{meu96} in another context).  The value of $\alpha $, of order of magnitude unity,   depends on the impact  geometry, the spacecraft and antenna geometry, and the size and charge of the impact plasma cloud, with respect to the shielding capabilities of both the ambient plasma and the photoelectrons ejected by the spacecraft and antennas.

If the two antenna arms are separated by more than a Debye length, the arm closer to the impact site experiences a much higher voltage than the other one, so that the signal can be measured both in monopole and dipole mode with a similar order of magnitude; otherwise, the dipole voltage will be somewhat smaller, by an amount depending on the  asymmetry of the impact with respect to the antennas. As noted above, the voltage sign agrees with the \textit{Wind} dipole observations. This induced voltage is of the same sign as  the voltage produced on a monopole antenna by charge recollection by the spacecraft, but  generally much smaller (by the factor $ \alpha C_{sc}/4\pi \epsilon_0 L$), so that it may be barely seen with a monopole antenna. However, it may be the dominant detecting mechanism with thin dipole antennas having  well-separated arms as in the case of \textit{Wind}, or in dense magnetospheres when photoelectron emission is negligible; in the latter case, since the non-recollected charges would be the cloud's electrons, possibly moving faster than those of a cold  ambient plasma, the calculation of the voltage is more complicated, possibly involving plasma waves.


Consider a grain of mass $m$ impacting at speed $v$, producing the impact charge $Q \simeq 0.7 m_{\rm{kg}}^{1.02} v_{\rm{km/s}}^{3.48} $ C  \citep{mcb99}. The above estimate yields a  pulse of  peak maximum amplitude
\begin{equation}
\delta V \sim  0.7 m_{\rm{kg}}^{1.02} v_{\rm{km/s}}^{3.48} \Gamma \alpha /(4\pi \epsilon _0 L) \; \rm{Volts} \label{V}
\end{equation}
on an antenna arm of length $L$ (in m) and receiver gain $\Gamma$.

With  a 10 nm radius nanograin of mass $m \simeq 10^{-20}$ kg impacting at 300 km/s, as predicted by dynamics \citep{man14}, $L \simeq 7.5 $ m and $\Gamma \simeq 0.4 $ for the \textit{Wind}/WAVES $E_y$ dipole \citep{sit97}, Eq.(\ref{V}) yields a voltage pulse in the mV range -  below the sensitivity of the instrument, and a still smaller voltage on the longer $E_x$ dipole.


In contrast, a  grain of  radius 1 $\mu $m ($m \simeq 10^{-14}$ kg) impacting at about 30 km/s should yield  from (\ref{V}) a voltage pulse on the \textit{Wind} dipole antennas of order of magnitude 200 mV,  which suggests that this mechanism is able to explain the range of amplitudes observed.

\section{Conclusions}

We conclude that:
\begin{itemize}
\item  our analysis taking into account the waveform capture capability of \textit{Wind}/WAVES in dipole mode indicates that this instrument is unable to detect the interplanetary nanodust discovered by STEREO;
\item  a new dust detection mechanism is proposed which should  explain the amplitude and polarity of the voltage pulses observed on \textit{Wind} when  dust grains in the tenths of micron to micron size range impact the spacecraft;
\item  this new mechanism may be the dominant dust detection mechanism by wave instruments using thin dipole antennas on other spacecraft; in particular, its applicability to the dipole observations on Voyager, Cassini and future spacecraft instrumentation, in addition to \textit{Wind}, should be examined in detail.
\end{itemize}

%
%

\begin{acknowledgments}
The Cassini/RPWS HF receiver data are available at University of Iowa and at LESIA (Observatoire de Paris, France). We thank the CNES and the CNRS for funding.
\end{acknowledgments}

\bibliographystyle{agu08} \bibliography{Biblio_total_V8}

\begin{thebibliography}{}

\bibitem[\textit{Aubier et al.}(1983)]{aub83} Aubier, M. G., N. Meyer-Vernet, and B. M. Pedersen (1983), Shot noise and particle impacts in Saturn's ring plane, \textit{Geophys. Res. Lett.}, 10, 5-8.
\bibitem[\textit{Belheouane et al.}(2012)]{bel12}Belheouane S. et al. (2012), Detection of interstellar dust with STEREO/WAVES at 1 AU, \textit{Solar Phys.}, 281, 501–506.

\bibitem[\textit{Graps et al.}(2001)]{gra01}Graps, A. L. et al.  (2001), Io revealed in the Jovian Dust Streams, in \textit{Proceed. Meteor. 2001 Conf.},  ESA SP-495, edited by B. Warmbein, pp. 601-608, ESTEC, Nordwijk.
\bibitem[\textit{Gurnett et al.}(1983)]{gur83} Gurnett, D.A. et al. (1983), Micron-sized particles detected near Saturn by the Voyager plasma wave instrument, \textit{Icarus}, 53, 236–254.

\bibitem[\textit{Gurnett et al.}(1986)]{gur86} Gurnett, D.A, T. F. Averkamp, F. L. Scarf and E. Gr\"{u}n  (1986), Dust particles detected near Giacobini-Zinner by the ICE Plasma Wave instrument, \textit{Geophys. Res. Lett.}, 13, 291–294.
\bibitem[\textit{Gurnett et al.}(1987)]{gur87} Gurnett, D.A et al.  (1987), Micro-sized particle impacts detected near Uranus by the Voyager 2 Plasma Wave instrument, \textit{J. Geophys. Res.}, 92, 14959-14968.

\bibitem[\textit{Gurnett et al.}(1991)]{gur91} Gurnett, D.A, W. S. Kurth, L. J. Granroth and S. C. Allendorf   (1991), Micro-sized particles detected near Neptune by the Voyager 2 Plasma Wave instrument, \textit{J. Geophys. Res.}, 96, 19,177–19,186.


\bibitem[\textit{Gurnett et al.}(1997)]{gur97} Gurnett, D.A, J. A. Ansher, W. S. Kurth and L. J. Granroth,   (1997), Micron-sized dust particles detected in the outer solar system by the Voyager 1 and 2 plasma wave instruments, \textit{Geophys. Res. Lett.}, 24, 3125–3128.


\bibitem[\textit{Gurnett et al.}(2004)]{gur04}Gurnett, D. A. et al. (2004), The Cassini radio and plasma wave investigation, \textit{Space Sci. Rev.}, 114, 395-463.
\bibitem[\textit{Kurth et al.}(2006)]{kur06}Kurth, W. S., T. F. Averkamp, D. A. Gurnett, and Z. Wang (2006),  Cassini RPWS observations of dust in Saturn's E ring, \textit{Planet. Space Sci.}, 54, 988-998.
\bibitem[\textit{Le Chat et al.}(2013)]{lec13}Le Chat G. et al. (2013), Interplanetary Nanodust Detection by the Solar Terrestrial Relations Observatory/WAVES Low Frequency Receiver, \textit{Solar Phys.}, 286, 549–559.
\bibitem[\textit{Malaspina et al.}(2014)]{mal14}Malaspina D. et al. (2014),  Interplanetary and interstellar dust observed by the Wind/WAVES electric field instrument, \textit{Geophys. Res. Lett.},  41, 266-272.

\bibitem[\textit{Mann et al.}(2011)]{man11} Mann, I. et al. (2011), Dusty plasma effects in near Earth space and  interplanetary medium, \textit{Space Sci. Rev.}, 161, 1-47. 
\bibitem[\textit{Mann et al.}(2014)]{man14} Mann, I., N. Meyer-Vernet and A. Czechowski, (2014), Dust in the planetary system: Dust interactions in space plasmas of the solar system, \textit{Physics Reports}, 536, 1-39.
\bibitem[\textit{McBride   \& McDonnell}(1999)]{mcb99} McBride, N.  and  J. A. M.  McDonnell (1999), Meteoroid impacts on spacecraft: sporadics, streams, and the 1999 Leonids, \textit{Planet. Space Sci.}, 47, 1005-1013.
\bibitem[\textit{Meuris et al.}(1996)]{meu96}Meuris, P., N. Meyer-Vernet, and J. F. Lemaire    (1996), The detection of dust grains by a wire dipole antenna: the radio dust analyzer, \textit{J. Geophys. Res.}, 101, 24,471-24,477.
\bibitem[\textit{Meyer-Vernet}(1985)]{mey85} Meyer-Vernet, N. (1985), Comet Giacobini-Zinner diagnosis from radio measurements, \textit{Adv. Space Res.} 5, 37-46.

\bibitem[\textit{Meyer-Vernet et al.}(1986a)]{mey86a} Meyer-Vernet, N., M. G. Aubier and B. M. Pedersen   (1986a), Voyager 2 at Uranus: Grain impacts in the ring plane, \textit{Geophys. Res. Lett.}, 13, 617–620.

\bibitem[\textit{Meyer-Vernet et al.}(1986b)]{mey86b} Meyer-Vernet, N. et al. (1986b), Plasma diagnosis from thermal noise and limits on dust flux or mass in Comet Giacobini-Zinner, \textit{Science}, 232, 370–374.

\bibitem[\textit{Meyer-Vernet  and Perche}(1989)]{mey89}Meyer-Vernet, N. and C. Perche (1989) Toolkit for antennae and thermal noise near the plasma frequency, \textit{J. Geophys. Res.}, 94, 2405-2415.

\bibitem[\textit{Meyer-Vernet  et al.}(1996)]{mey96} N. Meyer-Vernet, A. Lecacheux and B. M. Pedersen,  (1996), Constraints on Saturn's E ring from the Voyager-1  radioastronomy instrument, \textit{Icarus}, 123, 113-128.

\bibitem[\textit{Meyer-Vernet  et al.}(1998)]{mey98} N. Meyer-Vernet, A. Lecacheux and B. M. Pedersen,  (1998), Constraints on Saturn's G ring from the Voyager-2  radioastronomy instrument, \textit{Icarus}, 132, 311-320.


\bibitem[\textit{Meyer-Vernet}(2001)]{mey01} Meyer-Vernet, N. (2001), Detecting dust grains with electric sensors: planetary rings, comets, and the interplanetary medium, in   \textit{Proc. 7th Spacecraft Charging Technology Conf.}, ESA SP-476, edited by R. A. Harris, pp. 635-640, ESTEC, Nordwijk.

\bibitem[\textit{Meyer-Vernet et al.}(2009a)]{mey09a}Meyer-Vernet, N., et al. (2009a), Voltage pulses on STEREO/WAVES: nanoparticles picked-up by the solar wind?, \textit{Solar Phys.}, 256, 463-474.
\bibitem[\textit{Meyer-Vernet et al.}(2009b)]{mey09b}Meyer-Vernet, N., A.  Lecacheux,  M.L. Kaiser, and D.A. Gurnett (2009b), Detecting nanoparticles at radio frequencies: Jovian dust stream impacts on Cassini/RPWS, \textit{Geophys. Res. Lett.}, 36, L03103.
\bibitem[\textit{Meyer-Vernet and Zaslavsky}(2012)]{mey12}Meyer-Vernet, N. and A. Zaslavsky (2012) 
In Situ Detection of Interplanetary and Jovian Nanodust with Radio and Plasma Wave Instruments, in \textit{Nanodust in the solar system: discoveries and interpretations}  edited by I. Mann et al., Springer,  pp. 133-160.
\bibitem[\textit{Moncuquet and Schippers}(2013)]{mon13}Moncuquet, M.  and P. Schippers,   Eos Trans. AGU, XX, Fall
  Meet. Suppl., Abstract P32B-05.
\bibitem[\textit{Oberc}(1996)]{obe96}Oberc  P (1996), Electric antenna as a dust detector, \textit{Adv. Space Res.},  17,   (12)105-(12)110
\bibitem[\textit{Pantellini et al.}(2012)]{pan12}Pantellini  F, S. Belheouane, N. Meyer-Vernet, and A. Zaslavsky  (2012), Nano dust impacts on spacecraft and boom antenna charging, \textit{Astrophys. Space Sci.},  341,   309--314
\bibitem[\textit{Pantellini et al.}(2013)]{pan13}Pantellini  F et al. (2013), On the detection of nano dust using spacecraft based boom antennas, \textit{Solar Wind 13} ({\it AIP Conference Proceedings\/} vol 1539) ed~G P~Zank et al. (American Institute of Physics)  pp. 414-417.

\bibitem[\textit{Pedersen et al.}(1991)]{ped91}Pedersen, B. M. et al.  (1991), Dust distribution around Neptune: Grain impacts near the ring plane measured by the Voyager radioastronomy experiment, \textit{J. Geophys. Res.},  96, 19,187–19,196.


\bibitem[\textit{Schippers et al.}(2013)]{sch13}Schippers P., M. Moncuquet, N. Meyer-Vernet and A. Lecacheux  (2013), Core electron temperature and density in the innermost Saturn’s magnetosphere from HF power spectra analysis on Cassini, \textit{J. Geophys. Res.},  118, 7170–7180.

\bibitem[\textit{Sitruk and Manning}(1997)]{sit97}Sitruk, L. et R. Manning    (1997), L'experience spatiale G.G.S./\textit{Wind}/WAVES: ondes radioélectriques et ondes de plasma, CDPP Technical Document.

\bibitem[\textit{Zaslavsky et al.}(2012)]{zas12}Zaslavsky, A. et al.    (2012), Interplanetary dust detection by radio antennas: mass calibration and fluxes measured by STEREO/WAVES, \textit{J. Geophys. Res.}, 117, A05102.

\end{thebibliography}

%
%
%

\end{article}




%

%

%

\end{document}